# Role of local mode mixity and stress triaxiality in the fracture of niobium/alumina bi-crystal interfaces – a CPFEM based study

A. Siddiq[a], S. Schmauder[b]
[a] School of Engineering, University of Aberdeen, Aberdeen, UK
[b] Institute for Materials Testing, Materials Science and Strength of Materials (IMWF), University of Stuttgart, Germany

**Abstract**
Local mode mixity and stress triaxiality plays an important role during metal/ceramic interface fracture. It has been reported in the literature that the ratio of the fracture energies to the mode I fracture energy can vary from 1 to 5. For many of the bimaterial systems, the reason has been reported to be the local mode mixity and stress triaxiality. Korn et al. [1] performed experimental studies on bicrystal niobium/alumina interfaces and for some orientations the values of the above ratio were reported to be more than 5. However, it was believed that local mode mixity and stress triaxiality might not be the reason for this. In the presented work an effort has been put to explain the role of local mode mixity and stress triaxiality in the fracture of niobium/alumina bi-crystal interfaces using crystal plasticity based constitutive model.

**1. Introduction**
Metal/ceramic joints have become more and more important in modern technology, because they combine the properties of metals like ductility, high electrical and thermal conductivity and the properties of ceramics like high hardness, corrosion resistance and capacity of resistance to wear. The fracture at or near metal/ceramic interfaces often limits the reliability of these joints. Therefore, knowledge of the stress and deformation fields at the crack tip of a metal/ceramic interface is needed in order to develop a fundamental understanding of the fracture process. In many situations, cracks initiate at interfaces and advance along, towards or away from the interfaces.

Authors have presented in the past a multiscale length scale bridging approach to simulate the niobium/alumina bicrystal interface fracture [2-6]. Results were presented using local [2-4, 6] and nonlocal [5] crystal plasticity theories and explained micromechanics based physical mechanisms (crystal orientation, slip system activation, etc.) involved during fracture. A relationship among macroscale fracture energy, microscale plasticity and nanoscale (sub-micron scale) fracture parameters was presented after performing rigorous computational studies [2, 6].

In the past it has been reported that significant changes in the locus of failure (fracture energies) and the crack propagation behaviour could occur due to the local stress state of bimaterial interfaces [7-16]. In the present work, an effort has been made to understand the role of local mode mixity and stress triaxiality during niobium/alumina bicrystal interface fracture. It is aimed to understand how local mode mixity and stress triaxiality affect the macroscale fracture energies, locus of failure and crack propagation behaviour for a stationary and growing crack.

## 2. Computational Model

Details of the finite element model of bicrystal specimen, crystal plasticity based constitutive model for single crystal niobium, and cohesive model for interface fracture have been discussed in detail in the past (see ref [2, 5, 6]). However, a brief description of the computational model has been presented in the following for clarity. The dimensions of the finite element model were 2 x 4 x 36 mm$^3$ with a notch length of 0.4 mm (Figure 1).

The specimen is loaded with a loading rate of 96.8 micron/min. Both outer alumina shanks and the alumina single crystalline material, at the centre of the specimen, were treated as purely elastic. The polycrystalline niobium sheet was always modelled using Ramberg-Osgood relation. Single crystalline niobium is always modelled using BCC single crystal plasticity theory with three stage analytical hardening model by Bassani and Wu [17]. The hardening expression consists of self and latent hardening as a function of the plastic slip in all slip systems. Material parameters used were identified using inverse modelling technique (please see reference [21])

For crack propagation analyses along the interface, a cohesive model proposed by Scheider and Brocks [18] was used. Simulations were performed for three orientations of niobium/alumina bicrystal interfaces, i.e. Orientation I (niobium(110)[001]|alumina(11-20)[0001]), Orientation II (niobium(100)[001]|alumina(11-20)[0001]); and Orientation III (niobium(111)[-1-12]|alumina(11-20)[0001]).

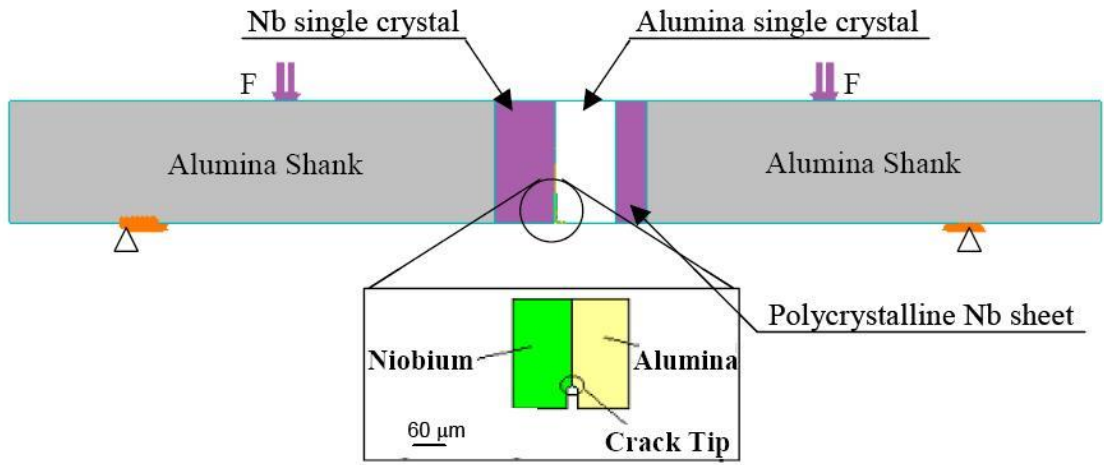

Figure 1: Schematic of Four-Point-Bending-Test-Specimen [2, 5, 6]

## 2. Local mode mixity at the crack tip

The local mode mixity at the crack tip is defined by the local phase angle $\psi'$. The variation in the local phase angle $\psi'$ may change the response of the crack tip upon loading on a bonded interface in metal/ceramic biomaterial systems. The local phase angle $\psi'$ describes the model mixity at the crack tip and is not necessary equal to the external loading phase angle and this difference comes from the mismatch in elastic and thermal properties of the materials involved.

This local phase angle $\psi'$ and the energy release rate $G_c(\psi')$ when plotted, give the failure locus as shown below in Figure 2. O'Dowd et al. [15] and O'Dowd [7] discussed the effect of local mode mixity on the fracture energies of a niobium/alumina bimaterial specimen. They found the same behaviour as Wang and Suo [19] between local phase angle and stress intensity factors as shown in Figure 2.

Shih and Asaro [8] showed that the local phase angle $\psi'$ for the case of elastic plastic interface fracture can be computed via the following relation:

$$\psi' = arctan\left(\frac{\sigma_{xy}}{\sigma_{xx}}\right)$$

where $\sigma_{xy}$ denotes the shear stress at the crack tip along the crack front, while $\sigma_{xx}$ is the normal opening stress at the crack tip at a distance of L=r along the crack propagation direction. Using above relation the local phase angle $\psi'$ has been computed for the case of a stationary crack tip and for growing cracks.

The local phase angle $\psi'$ for the case of a stationary crack tip in bicrystal niobium/alumina system for three different orientations is plotted in figure Figure 2. Experimental locus of failure [7] is also plotted as solid line. It can be inferred from figure Figure 2 that as the shear stress values along the interface at the stationary crack tip are small compared to the normal opening stress, the local phase angle $\psi'$ is in the range of -4.0 to -3.0 degrees. This range of local phase angles $\psi'$ shows that the effect of local mode mixity is almost negligible. This can also be seen from figure which shows for this range of local phase angles $\psi'$ the $K_c/K_{Ic}$ ratio is approximately equal to 1.

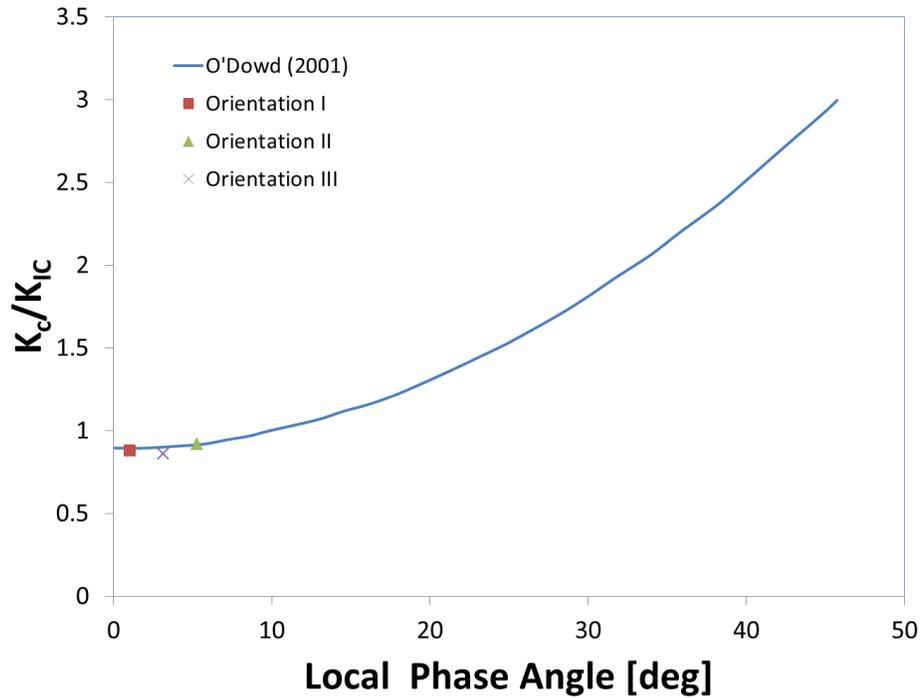

Figure 2: Comparison of experimental fracture toughness curve [7] with presently computed local phase angle through crystal plasticity based simulation for three different orientations

Similarly, local phase angles $\psi'$ have also been computed for growing cracks using a cohesive modelling approach and are plotted in Figure 3. Two different cases have been considered. The first one is when both, the normal cohesive strength and the shear cohesive strength values are selected to be equal. It must be noted that the normal cohesive strength is the maximum normal stress required for damage initiation and the shear cohesive strength is the maximum shear stress required damage initiation. The second case is when the shear cohesive strength is selected as half of the normal cohesive strength.

The local phase angle $\psi'$ for the case both the cases at different crack tip positions during crack growth are plotted in Figure 3. All of the results are for orientation II. The values of the local phase angle $\psi'$ are found to be in the range of -7.0 to -4 degrees.

The result of the local phase angle $\psi'$ for the case of a stationary crack tip and for growing cracks show that the effect of local mode mixity is almost negligible for the case of niobium/alumina bicrystal specimen. Therefore, local mode mixity does not play any significant role during niobium/alumina bicrystal specimen.

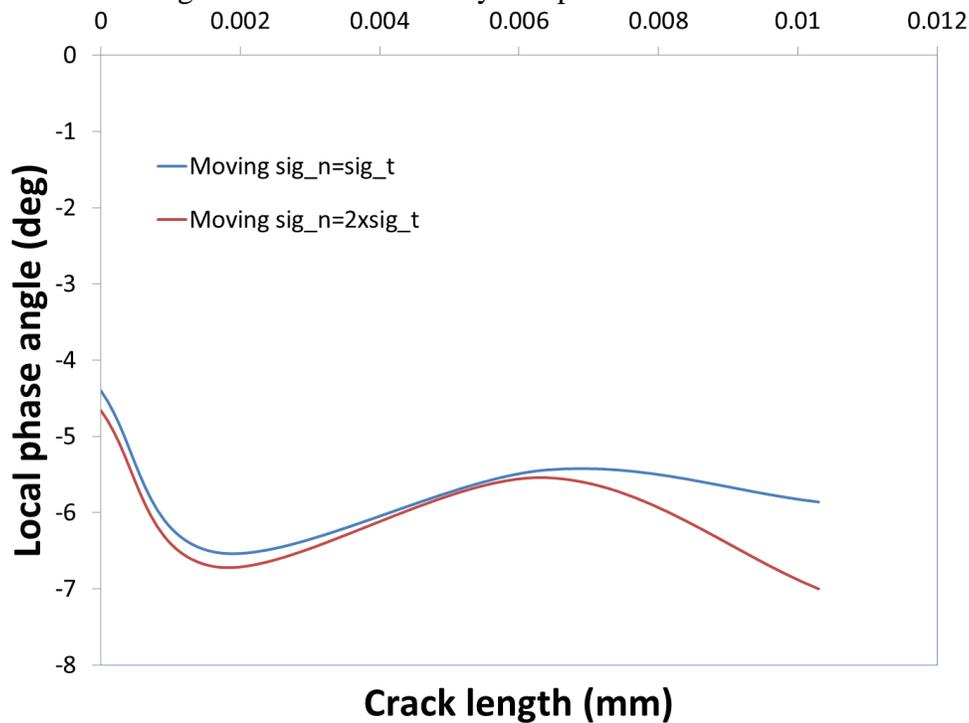

Figure 3: Computed local phase angle for orientation II as crack grows

## 3. Stress Triaxiality at the Crack Tip along the Interface

The locus of failure and the crack propagation behaviour are significant aspects in evaluating the mechanical properties of bonded joints. Dillard [9] showed that the crack path locus is closely related to material properties such as strength, interface quality and fracture toughness of the bonds along with the stress state at the crack tip.

Cao and Evans [10] and Akisanya and Fleck [11] showed that the locus of failure and the crack propagation behaviour are dependent on the mode mixity of external loads. Cotterell and Rice [14] showed that the T-stress (stress acting in the direction parallel to the crack plane) plays an important role in the directional stability of crack propagation. The crack is directionally stable if the T-stress is negative, whereas it is directionally unstable if the T-stress is positive. The reason for the directional instability in case of positive T-stress is based on the fact that higher the positive T-stress, higher will be the stress triaxiality causing the crack to kink towards a direction where the T-stress becomes negative [9, 14]. It must be noted that a negative T-stress means compressive stress which does not allow the crack to kink while the positive T-stress means tensile stress which makes a crack to kink away from a direction where the T-stress is compressive. Rice [14] showed this crack kinking behaviour analytically (see [14] for details), it was shown that the crack kinks towards a direction where the T-stress becomes negative which is also the same direction where $K_{II} = 0$. Fleck et al. [12] and Akisanya and Fleck [11, 13] also concluded that under predominantly mode I loading, the crack propagation in an adhesive bond is directionally stable if the T-stress is negative and is directionally unstable if the T-stress is positive.

This criterion, although developed primarily for cracks in homogeneous materials, can be readily extended to interfaces in bimaterial systems such as bonded joints according to Hutchinson [16].

The study of Zhu and Chao [20] provides further insights about the effect of T-stress on the crack propagation behaviour in homogeneous media. Not only the directional stability of cracks, but also the direction of crack propagation will be affected by the T-stress. Zhu and Chao [20] clarified that although the criteria for cracking direction and direction stability of cracks are developed under the assumptions of linear elastic fracture mechanics, they are still applicable for ductile materials.

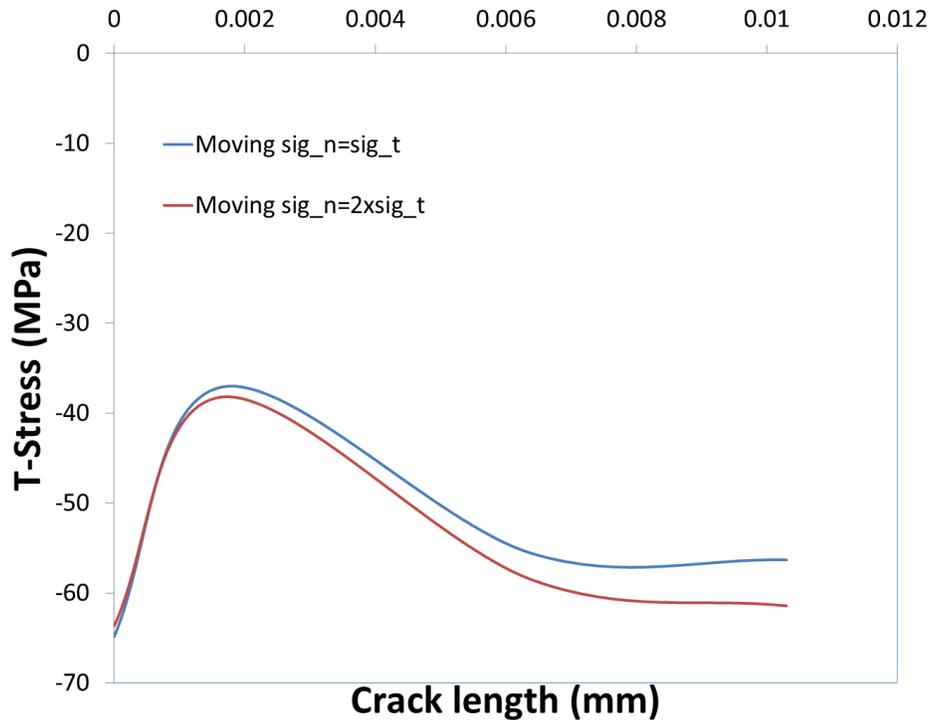

Figure 4: Computed T-stress for orientation II as crack grows

For pure mode I fracture, i.e. $K_{II} = 0$, the T-stress along the crack plane ($\theta = 0$ and $\pm\pi$) is given by

$$T = \sigma_{xx} - \sigma_{yy}$$

If the fracture is in mixed mode, i.e. $K_{II} \neq 0$, then T-stress behind the crack tip ($\theta = \pm\pi$) is given by [9]

$$T = \sigma_{xy} + K_{II}\sqrt{\frac{2}{\pi r}}$$

with

$$K_{II} = \left(\sigma_{xy}\sqrt{2\pi r}\right)_{\theta=0}$$

The relations for T-stress discussed above are used to compute the T-stress values at various crack tip positions during crack growth for two different combinations of normal cohesive strength and shear cohesive strength.

The computed values of T-stress for various crack lengths are plotted in Figure 4. The results show that T-stress are always found to be negative, therefore, the crack propagation directions is always stable according to the above criteria, i.e. along the interface during all the simulations.

## 4. Conclusion

Role of local mode mixity and stress triaxiality at the crack tip was discussed for the case of niobium/alumina bicrystal interface fracture. It was found that for short cracks the local phase angle is always in the range of -7 to -4 degrees, which when plotted on fracture toughness curve of niobium/alumina interface shows that the influence of local mode mixity is almost negligible.

The effect of stress triaxiality was also studied to find the directional stability of the growing cracks. T-stresses were computed and are always found to be negative in the range of -65 to -35 MPa resulting in a directionally stable crack growth, namely along the interface of niobium/alumina bicrystal specimens because of low stress triaxiality at the interface crack tip of the niobium/alumina bicrystal interface. The physical reason as explained above is negative T-stress means compressive stress acting parallel to the interface causing the crack to propagate in the direction along the interface while a positive T-stress means tensile stresses acting parallel to the interface causing an interface crack to kink into the direction where the stress parallel to the crack face becomes compressive.